\documentclass[aps,pre,twocolumn]{revtex4-1}
\usepackage{graphicx}
\begin{document}

%
%

\title{Critique of the Gibbs volume entropy and its implication}   

\author{Jian-Sheng Wang}
\email[]{phywjs@nus.edu.sg}
\affiliation{Department of Physics, National University of Singapore, Singapore 117551, Republic of Singapore}
\date{8 July 2015}

\begin{abstract}
Dunkel and Hilbert, ``Consistent thermostatistics forbids negative 
absolute temperatures,'' Nature Physics, {\bf 10}, 67 (2014), and 
Hilbert, H\"anggi, and Dunkel, ``Thermodynamic laws in isolated 
systems,'' Phys. Rev. E {\bf 90}, 062116 (2014) have presented an 
unusual view of thermodynamics  
that sets aside several properties that have traditionally have been assumed
to be true.
Among other features, their results do not satisfy  the postulates of 
thermodynamics
originally proposed by Tisza and Callen.
In their theory, differences in the temperatures of two objects  
cannot
determine the direction of heat flow when they are brought into thermal 
contact. They deny that negative temperatures are possible.
%
We disagree with most of their assertions, and we present arguments
in favor of a more traditional interpretation of thermodynamics.
We show  explicitly 
that it is possible to 
deduce the thermodynamic entropy for a paramagnet along the lines of
classical thermodynamics and that it agrees with Boltzmann
entropy in equilibrium.    
A Carnot engine with efficiency larger than one is a matter of definition 
and is possible for inverted energy distributions,
regardless of whether negative temperatures are 
used in the analysis.
We elaborate on Penrose's argument that an adiabatic and
reversible process connecting systems with Hamiltonian $H$ to $-H$ is possible, thus negative temperatures logically must exist.  We give a demonstration
that the Boltzmann temperature determines the direction of heat flow while Gibbs
temperature does not, for sufficiently large systems. 
\end{abstract}

\maketitle

\section{Introduction}
Compelling arguments have been presented for the thermodynamic concept of negative temperature, an idealization and abstraction for systems with bounded energies \cite{purcell,ramsey,schreider-exp}.   However, Dunkel and Hilbert (DH)   \cite{DH} 
and Hilbert, H\"anggi, and Dunkel (HHD) \cite{HHD}   
have recently argued  that negative temperatures arise due to the use of the Boltzmann entropy, which they claim to be inconsistent.  Instead, they suggest Boltzmann's entropy should be replaced by one due to Gibbs.  The Gibbs entropy is defined as the Boltzmann constant times the logarithm of the total phase space volume less than a given energy.   Since the phase space volume is an increasing function of energy, the Gibbs entropy only allows for positive temperatures. 
 
For quantum systems with discrete energy levels, the Boltzmann entropy corresponds to the logarithm of the degeneracy of an energy eigenstate 
(or in a narrow energy interval), while the Gibbs entropy is obtained by counting all the states less than or equal to a given energy \cite{DH,DHreply,DH reply to Schneider,HHD,campisi}.
With this definition of entropy, temperature cannot be negative as entropy is a
nondecreasing function of energy. While the differences
between the predictions of the Gibbs and Boltzmann entropies are unmeasurable 
for macroscopic systems with monotonic energy densities (such as a system of harmonic oscillators), the differences are dramatic for systems with non-monotonic density of states.


DH and HHD also applied 
 their theory of thermodynamics to
systems with very few degrees of freedom.
We do not believe that thermodynamics is applicable to systems 
for which the relative fluctuations are large,
so we will restrict our remarks to large systems.

Objections to DH's claims have been made by several authors \cite{schneider,vilar,frenkel,swendsen-wang2014,swendsen-wang2015,penrose-counter-eg,buonsante}.  In this paper, we expand on these 
discussion to emphasize 
that it is appropriate to include negative temperature in thermodynamics,
 and to
illustrate some incorrect
predictions of Gibbs entropy.


The organization of the paper is as follows: In Sec.~\ref{thermo-entropy}
we briefly comment on the formulation of thermodynamics and entropy. 
A  
consistency condition is discussed in Sec.~\ref{condition}, and we point out that
such a condition has no constraint to the Ising model. 
In 
Sec.~\ref{S-para}, we discuss the entropy of  a paramagnet.  
After that, in Sec.~\ref{carnot}, we discuss why an efficiency larger than one is 
possible for a Carnot engine for inverted energy distributions,
and that they are well described by 
negative temperatures.   
We then discuss the
properties of adiabatic invariance in classical (Sec.~\ref{adiabat-classical})
and quantum systems (Sec.~\ref{SecQadiabaticinvarant}) and demonstrate
that the number of states less than a given value of energy is
not a true adiabatic invariant.   We demonstrate with the examples of model
systems that negative temperature does not lead to contradictions in 
Sec.~\ref{secIsing} and \ref{sec-toy}.  In Sec.~\ref{one-degree}
we criticize the   
application of thermodynamics      
to systems with only a single one degree of freedom. 
 In Sec.~\ref{Sgproblem}, a two-level model 
calculation shows that Boltzmann temperature is decisively the
correct temperature 
since it predicts the direction of heat transfer
and is
consistent with the second law of thermodynamics. 
After briefly discussing ensemble equivalence (Sec.~\ref{sec-ensemble-eq})  
and the additivity of entropy (Sec.~\ref{sec-additive}), we discuss the
concavity of the Gibbs entropy (Sec.~\ref{secCV}) and its violation
of Callen's postulates (Sec.~\ref{sec-postulate-II}) for negative  
temperature states. We discuss the thermodynamics formulated 
in Ref.~\onlinecite{HHD} (Sec.~\ref{sec-HHD-ThermoD}) and considering
the nonequilibrium statistical mechanics of 
splitting and joining of two systems (Sec.~\ref{sec-split-join}).  
We remark on the non-monotonic dependence of temperature with energy
for finite systems and why it disappears (Sec.~\ref{sec-non-mono}) for large
systems (or in thermodynamic limit \cite{Nlimit}).
We conclude with some final remarks in Sec.~\ref{sec-why} and \ref{sec-final}.

\section{\label{thermo-entropy}Thermodynamics and Entropy}

 Dunkel and Hilbert
 \cite{DH} began with a statement that the
positivity of absolute temperature is a key postulate of thermodynamics.  
DH cited Callen (postulate III, page 28, 2nd edition,
\cite{callen}) that entropy `is a monotonically increasing function of the
energy.' 
They did not explain why they regard it as a key postulate.
The only reason we know of is 
that monotonicity allows the equation 
$S=S(E,V,N)$
to be inverted to give
$E=E(S,V,N)$,
from which the usual thermodynamic potentials can be obtained 
by Legendre transforms.
This is a matter of convenience, not principle,
and Massieu functions can produce the same results
through Legendre transforms of
$S=S(E,V,N)$.
Apparently, DH rejected most of Callen's other postulates of
thermodynamics.


The axiomatic formulation of thermodynamics, such as that given by Giles
\cite{Giles}, or Lieb and Yngvason \cite{Lieb}, does not feature the concept of
temperature until a very late stage.  So the positivity of temperature is not
at all a fundamental axiom of thermodynamics.  Even energy  does
not enter into the picture initially.  What is fundamental in thermodynamics is
the concept of irreversibility.  This concept appears and is an emergent
phenomenon only in the macroscopic world; as is well-known, there is no
irreversibility in microscopic physical laws. 

Whether one follows the traditional approach to thermodynamics
\cite{pippard}, or the
postulatory approach of Gibbs, Tisza, and Callen, or the more precise axiomatic formulation in
the post-modern era, they are all consistent and convergent to the same
physical theory of thermodynamics.  When we say ``thermodynamics'', there
should not be disagreement as what we mean by it.  But since Hilbert,
H\"anggi, and Dunkel (HHD, \cite{HHD}) proposal of a thermodynamics
is substantially different from the usual formulation, we need to examine
its legitimacy.

It is possible to develop statistical mechanics in exactly the same way as in
the traditional thermodynamics, where an empirical temperature occupies a
prominent place in defining isothermal processes. With the help of the Carnot cycle, we
can define absolute temperature, and thus, following Clausius, entropy.  This
chain of reasoning puts entropy at the last, so the dispute over which
expression is the correct entropy can be, hopefully, resolved.  We give some details with
such an approach here.  Of course, the final result is exactly the same as the
standard textbook statistical mechanics with Boltzmann entropy.

\subsection{Stability of thermodynamic systems}
DH and HHD have argued 
that negative temperature states are unstable or unphysical;
any contact with a positive temperature environment will collapse the system 
\cite{romero}.  There seems to be a misunderstanding of thermodynamic
stability. 
Any  thermal contact of any system with another system at a different temperature
takes it through an irreversible process into a different equilibrium state.
If a system at negative temperature is brought into contact 
with another system at negative temperature,
the final equilibrium will be at an intermediate tempearature
that is also negative.
The states are stable 
if the entropy associated with the system is a concave function of
energy.
If a negative temperature
system is in contact with a system of the same negative temperature, it will
happily coexist.    

%


\subsection{Consequences for the definition of entropy}

We choose a notation that is largely consistent with that of DH.  The total number
of states below energy $E$ is
\begin{equation}
\Omega(E) = \sum_{E_j \leq E} 1.
\end{equation}
For simplicity, we'll consider mostly quantum systems with a 
set of discrete energy eigen spectra, $E_j$, $j=1$, 2, $\cdots$.  And
$\omega(E)\Delta \equiv \Omega(E) - \Omega(E-\Delta) \approx
\Omega'(E) \Delta$.  The Gibbs (volume) entropy is $S_G = k_B \ln \Omega(E)$,
and Boltzmann,  
$S_B = k_B \ln (\omega(E)\Delta)$.  The small energy
$\Delta$ is used to specify an energy range of microcanonical ensemble of 
the system.  

%
%
%

We define
$T^{-1}_G = \partial S_G / \partial E$ and
$T^{-1}_B = \partial S_B / \partial E$.

We had always thought that the difference between $S_B$ and $S_G$ was so small
for a macroscopic system that it would be impossible to detect, which would
imply that the choice might be a matter of taste to some extent. 
This is true for systems of particles with unbounded energies.
Now it
appears that DH have (inadvertently) provided a demonstration that $S_G$ is not
tenable.

%

In Refs.~\cite{swendsen-wang2014,swendsen-wang2015,RHS1,RHS book,RHS unnormalized,RHS Change
2014}, the definition of entropy, entropy of macroscopic observables, is based on the joint probability distribution
for large, but finite systems.  In this particular case, it is equivalent to
what DH labelled as $S_B$, although the additive constant is expressed
differently than their $k_B \ln \Delta$.  Therefore, by construction, the
maximum of the sum of the entropies occurs at the maximum probability of a subsystem energy
after the two subsystems have come to thermal equilibrium.

This means that -- if they insist on taking the small differences seriously --
the maximum of their sum of entropies occurs at a non-equilibrium value of the energy
of the subsystems,
see Sec.~\ref{SecNtest}.
If they were to give initial conditions such that the two subsystems started at
the maximum of their total entropy, releasing the constraint (allowing heat transfer), then 
splitting again would lower the entropy, in violation of the second law \cite{swendsen-wang2014}. 

%
%
%

\section{\label{condition}Consistent conditions for entropy}
We would certainly agree with the importance of DH's Eq.(4),
but it would be more appropriate to write it as 
\begin{equation}
dS
= \frac{1}{T} dE
+
\frac{1}{T} \sum_j a_j dA_j,
\end{equation}
where
\begin{equation}
 \frac{1}{T}  =  \left(\frac{ \partial S }{ \partial E } \right)_{ \{ A_j \} },
\end{equation}
\begin{equation}
\frac{1}{T} a_j 
=
 \left(\frac{ \partial S }{ \partial A_j } \right)_{E, \{ A_k \} k \ne j }.
\end{equation}
The volume $V$ should be included in the $A_j$'s.  It would not be an
appropriate variable for most systems with bounded energy.
%


For classical Hamiltonian systems, 
the argument that $T_B \ne T_G$ seems unfounded because the difference is of the
order of $1/N$, such as for ideal gas, while fluctuations are of the order $1/\sqrt{N}$.  If
fluctuations are negligible, which is usually assumed in thermodynamics, then
$T_B-T_G$ is negligible.

The consistency issues have been addressed in the past.   Boltzmann himself
discussed it in what Boltzmann called `heat theorems', asking questions
of consistency with thermodynamics (see appendix 9A1 of Gallavotti's
textbook \cite{gallavotti}; see also the discussion in S.-K. Ma, Chap.~23
\cite{skma}).  It is not essential that $S_G$ satisfies the condition 
for all $N$, but for none with $S_B$, since thermodynamics is well-defined only
in the limit $N \to \infty$.  And in that limit, it does satisfy (Frenkel and
Warren \cite{frenkel}), so nothing is violated.


A key feature of the consistent condition, Eq.~(6) in DH which we write down
again here, 
\begin{equation}
\label{eqconsistent}
a_j = T \left({ \partial S \over \partial A_{j} }\right)_E = - \left\langle { \partial H \over \partial A_{j} } \right\rangle,
\end{equation}
which the authors do not seem to notice is that when it applies to, say, the
one-dimensional (1D) ferromagnetic Ising model,
\begin{equation}
H = - J \sum_{i=1}^N \sigma_i \sigma_{i+1}, \quad \sigma_{N+1}=\sigma_1,\quad \sigma_i = \pm 1,
\end{equation}
it just produces a ``null statement'', as there is no $A$-variable to use to
form the equation.  The thermodynamic equation is simply $dE = TdS$.
Alternatively, one can use the coupling $J$ as one of the external control
thermodynamic variable, i.e., $A=J$.  Then the consistency condition only
produces an identity, $E= \langle H\rangle$, for any reasonable expression of
$S$.   This is because the $S$-dependence gets cancelled between $T$ and
$\partial S/\partial J$, since $S(E,J)\equiv S(E/J)$.  
The consistency condition has no use to judge which
entropy expression is correct, at least for the Ising chain.
In Sec.~\ref{seccampisi} below, we also show that the consistent condition is violated for a paramagnet if Gibbs volume entropy is used.
Extending the adiabatic invariant and consistency properties of Gibbs volume
entropy, valid for classical Hamiltonian systems with continuous energy
spectra, to all systems, 
seems to us an over-generalization,
and need further justification.

%
%
%
%

\section{\label{S-para}Entropy of a paramagnet from classical thermodynamics}

Which of the alternative starting points of definitions for entropy is 
correct?  We address this question in a different perspective.
Instead of taking a postulatory approach to the expression for entropy,  we
derive it, following the classical approach to thermodynamics \cite{pippard}.
This consists of the following steps: 0) fundamental to
statistical mechanics is the equal-a-priori probability and microcanonical
ensemble.  This will  be our basis in the following consideration. In order to
reproduce thermodynamics from statistical mechanics, we work in the
thermodynamic limit (i.e., large system sizes). 1) Build an empirical
thermometer to realize the zeroth law of thermodynamics. 2) Use the thermometer to
determine the equation of state and thus adiabatic and isothermal processes.  3)
Build Carnot cycles to find the universal function relating the empirical
temperature to the Kelvin scale. 4) Define the entropy according to Clausius.  This
chain of reasoning puts entropy at the last as an output of the more
mechanical arguments, instead of starting with the entropy postulates.   As a
by-product, the concept of negative temperature results naturally from the
derivative of entropy with respect to energy, if steps 1) to 4) are not 
erroneous.  To do this in a very general setting is obviously difficult.  Here we
are modest, considering only the simplest possible model systems.  In
particular, we take a one-dimensional (1D) Ising model as our bath or
thermometer, and a non-interaction paramagnet as the system (see also Frenkel and
Warren, Ref.~\onlinecite{frenkel}).  Our logical conclusion is that the usual textbook
entropy formula is sound and implies the existence of negative absolute
temperature for systems with bounded energy.


\subsection{Equal a priori probability and microcanonical ensemble}

We assume that the dynamics is such that energy is conserved.  The
probability distribution function (for classical systems) or density matrices
(for the quantum case) follows the Liouville or von Neumann equation.  However,
our Ising model does not have an intrinsic dynamics; we assume
that the system follows some form of stochastic dynamics governed by a master
equation (Penrose \cite{penrose}).   We do not need to explicitly specify the dynamics as
we are not interested in the actual time-dependent processes, in other words,
we work in the domain of equilibrium statistical mechanics and quasi-static
processes.  The only thing that is required is that the system evolves in such
a way that micro-states with the same energy (or other
conserved quantities, e.g., magnetization) will have the same probability, i.e.,
the microcanonical assumption.  Another point is that we will work in large size
limit (i.e., thermodynamic limit).  Systems with few degrees of freedom in
isolation belong to the regime of mechanics and cannot possibly behave like a
thermodynamic system (e.g., cannot thermalize).  This helps in simplification of our derivations.


\subsection{1D Ising model as a thermometer}

We consider a 1D Ising chain with a periodic boundary condition without a
magnetic field term, with the Hamiltonian function given by
\begin{equation}
H_B = - J\sum_{i=1}^{N_B} \sigma_i^B \sigma_{i+1}^B,\quad  
\sigma_i^B = \pm 1,  \quad \sigma_{N_B+1}^B = \sigma_1^B.
\end{equation}
The super-/subscript $B$ indicates the degrees of freedom of the bath (or thermometer) in
contrast to the system without the sub- or superscripts.   Since we work in
microcanonical ensemble, the (statistical mechanical) system is uniquely
characterized by the total energy $E_B$ and number of spin $N_B$.  Now we
define an empirical temperature $\theta = \langle \sigma_1^B \sigma_2^B \rangle$.
The average $\langle \cdots \rangle$ is with respect to the microcanonical
ensemble.  It is clear that $E_B = -JN_B \theta$.  We note that
$-1 \leq \theta \leq
1$, so our thermometer readings are real numbers from $-1$ to 1, with negative
values corresponding to the usual negative temperature.  As we'll see later,
this temperature scale $\theta$ is essentially $\beta = 1/(k_B T)$ mapped to a
bounded interval. Unlike the ideal gas thermometer, our Ising thermometer can
measure both positive as well as negative temperatures. 

\subsection{\label{para-adiabats}Paramagnet working system}
We build a Carnot engine made of an Ising paramagnet, given by the
Hamiltonian
\begin{equation}
H =  - h \sum_{j=1}^N \sigma_j,\quad   \sigma_j = \pm 1.
\end{equation}
Let $N_{+}$ and $N_{-}$ be the number of spins with plus and minus signs,
respectively, then the energy is $E = -hN_{+} + h N_{-} = -h M$, and the total
magnetization is $M = N_{+} - N_{-}$.  A thermodynamic state is uniquely
specified by the energy $E$ and the magnetic field $h$, while one of the 
equation of state (magnetization) is $M = -E/h$.  In a microcanonical ensemble energy does not fluctuate.
By the first law of thermodynamics, $dE = \delta Q + \delta W$, we can identify
the heat and work as $\delta Q = - h\, dM$  and $\delta W = - M dh$,
respectively.  The work is related to the part of internal energy changeable
by the control parameter $h$. From the above, we note that adiabatic lines are very simple,
given by $M = {\rm const}$. -- horizontal lines in an $M$ v.s. $h$ diagram 
(see Sec.~\ref{SecQadiabaticinvarant} for further justification).
However, the isothermal curves are more difficult to specify, 
as we do not have a real equation of state
relating to the (empirical) temperature yet.

%
%
%

\subsection{Measuring the empirical temperature of the system}

To find the empirical temperature of the paramagnetic system, we take
initially two separate systems of the thermometer, characterized by the total
energy $E_B$, and the system, characterized by $E$ and $h$ (the sizes $N_B$ and
$N$ are also parameters, but they are considered as fixed constants).   We then
combine them into a new system with the Hamiltonian $H_{\rm tot} = H_B + H$.
As there is no explicit interaction term between the two subsystems in the
Hamiltonian and the dynamics is such that it makes $H_{\rm tot}$
microcanonical, the effect of combining them is to make each subsystem no
longer in microcanonical distribution but new marginal distributions dictated
by the overall microcanical distribution.  We now ask what the expectation
value  $\langle \sigma_1^B \sigma_2^B\rangle_{H_{\rm tot} = E_B + E}$ is.  If
this value is the same as before making the contact, $\theta$, then we say that
the system is at the empirical temperature $\theta$.  This defines the
self-consistent condition for the temperature of the paramagnetic system
\begin{equation}
\theta = \left< {1 \over N_B} \sum_{i=1}^{N_B} \sigma_i^B \sigma_{i+1}^B \right>_{H_{\rm tot}(\sigma^B, \sigma) = E_B + E}.
\end{equation}
This is an equation for $\theta$ as the right-hand side also contains $\theta$
due to $E_B = -JN_B \theta$. The result will depend on $E$ and $h$ in the
constraints.   The equation can be made more explicit as
\begin{equation}
\label{Eq(A)}
E_B = \bigl\langle e_B(\sigma^B) \bigr\rangle =  { \sum_i e_B(i)\, n^B(i)\, n(N_{-})  \over   \sum_i n^B(i)\, n(N_{-})} ,
\end{equation}
where the `density of states' (number of microscopic states at energy $e_B(i) = -JN_B +
2Ji$, $i$ even) of the 1D Ising is
\begin{equation}
n^B(i) = 2 {N_B! \over i! (N_B - i)!}, \quad   i = 0, 2, ..., N_B.
\end{equation}
We assume $N_B$ even for simplicity.  $i$ denotes the number of plus-minus
boundaries in a 1D Ising configuration.  Similarly, the density of states of the
paramagnet is
\begin{equation}
n(N_{-}) = n(N_{+}) = { N! \over N_{+}! N_{-}!}, \quad  N_{+} + N_{-} = N, 
\end{equation}
and $N_{+} = (N - E/h)/2$, and $N_{-} = (N+E/h)/2$.  
$i$ and $N_{-}$ are related by energy conservation
\begin{equation}
e_B(i) -(N-2N_{-}) h = E_B + E.
\end{equation}
Equation (\ref{Eq(A)})
would be difficult to evaluation in general but great simplification is
available in the thermodynamic limit, i.e., $N_B$ and $N$ are large (but still
finite quantities).  Then the summation is dominated by the largest weight,
$n^B n$.  We find the value $i$ for which the weight is a maximum.  Although
$i$ and $N_{+}$ or $N_{-}$ are integers, in the large size limit, we can treat them as
real numbers.  Using the Stirling approximation, 
$\ln j! = j \ln j - j$, for $j$ sufficiently large, and obtaining the derivative of 
$\ln (n^B n)$ with respect to $i$ and setting it to 0, we obtain the condition
ln $(i /(N_B -i))  = (J/h) \ln (N_{-}/N_{+})$.
In deriving this, we have assumed self-consistency on the right-hand side,
i.e., $E_B = -JN_B \theta$, $\theta = 1 - 2i/N_B.$ This optimal value of $i$ is
essentially our temperature of the system.  Instead of using integers $i$ and
$N_{+}, N_{-}$, we can replace them by more physical quantities, with $i$ by
$\theta$, and $N_{+}$, $N_{-}$ by $M$ and $N$, we find
\begin{equation}
\frac{h}{J} \ln \frac{1+\theta}{1-\theta}  = \ln \frac{N+M}{N-M}
\end{equation}
or 
\begin{equation}
\tilde \beta h = \frac{h}{J} \tanh^{-1}(\theta) = \tanh^{-1}(M/N),
\end{equation}
or $M = N \tanh(\tilde \beta h)$ with $\theta = \tanh(\tilde \beta J)$.  Reader
familiar with the canonical treatment of 1D Ising model and paramagnet will
immediately recognize that these results are identical to the usual formulas
in canonical ensemble if we identify $\tilde \beta$ with the usual $\beta =
1/(k_B T)$  (We do not use canonical ensemble because that would
presume ensemble equivalence and defeat the purpose of introducing
empirical temperature).
Here $\tilde \beta $ is a derived concept, characterizing, in an
alternative way, the temperature of the system or bath.  This simple picture
emerges only if thermodynamic limit is taken.  For notational simplicity,
we'll drop the tilde below but $\beta$ should be understood as constructed
above.

\begin{figure}
\includegraphics[width=\columnwidth]{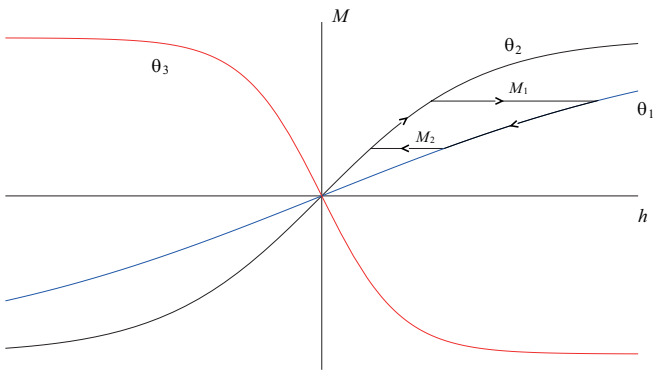}%
\caption{\label{figparamagnet}Carnot cycle using paramagnet as substance.}
\end{figure}


\subsection{Carnot cycle}
With the empirical temperature of the system well-defined, we can construct
Carnot cycles.  An example is given in Figure~\ref{figparamagnet}.   The two
adiabatic curves are characterized by $M_1$ and $M_2$ as horizontal lines.  The
isothermal curves are labeled by $\theta_1$ and $\theta_2$ (or equivalently
$\tilde\beta_1$ and $\tilde\beta_2$).  They are given by the equation of state
for the paramagnet as
\begin{equation}
M = N \tanh(\beta h).
\end{equation}
An important point is that these family of isothermal curves are identical no matter what
empirical temperature scale one uses, thus independent of the temperature
measuring devices.  Hence, the concept of `temperature' is just a one-parameter
family of real numbers labeling these curves.  Positivity of the temperature is
not required.  As in the usual construction of the Kelvin temperature scale, we
compute the heat absorbed at the high temperature, $Q_1$ at $\theta_1$,
and heat released at low temperature, $Q_2$ at $\theta_2$  with the total work,
$W > 0$, done by the system as the area of the loop running clockwise
on the cycle.  High and low temperatures are defined according to second law of
thermodynamics (not the actually values of the parameter $\theta$) - heat flows
from high temperature to low temperature if there is no external intervention.
Then,
\begin{equation}
{ Q_2 \over Q_1 } = { f(\theta_2) \over f(\theta_1) }.
\end{equation}
Our job is to find the function $f$, that depends on our choice of the
empirical temperatures, but if we define $T = {\rm const} f(\theta)$, then this
new temperature scale, $T$, the Kelvin scale, clearly is independent of the
makeup of the thermometer and is universal in the sense that all Carnot cycles
have exactly the same efficiency formula $\eta = W/Q_1 = 1 - T_2/T_1$ if temperatures
are the same on the Kelvin scale.  Thus, the Kelvin scale is a privileged one
over the empirical ones.  We now work out explicitly this function $f$.  Using
the formula for heat, we have, with the equation of state, 
\begin{eqnarray}
Q_1 &=& - \int_{M_1}^{M_2} h\, dM = -hM\Big|_{M_1}^{M_2} + \int_{M_1}^{M_2} M dh \nonumber \\
&=&   -h_2 M_2 + h_1 M_1 + { N \over \beta_1 } \ln  
{\cosh(\beta_1h_2) \over \cosh(\beta_1 h_1)} \nonumber \\
&=&  T_1 [S(M_2) - S(M_1)],
\end{eqnarray}
where we define $T_1 = 1/(k_B \beta_1)$, which is just the conventional Kelvin
scale, fixing an arbitrary constant relating our empirical temperature $\theta$
to $T$, and $S$ is a function defined by
\begin{eqnarray}
\label{Eq(B)}
S(M) &=& -k_B \beta h M + k_B N \ln \cosh(\beta h) + {\rm const} \nonumber \\
  &=& -  k_B N \Big[ \frac{1+M/N}{2} \ln \frac{1+M/N}{2} \nonumber \\ 
&&+ \frac{1- M/N}{2} \ln \frac{1-M/N}{2}\Big] .
\end{eqnarray}
To obtain the second line, we have used again the equation of state of the
system.  We also dropped an arbitrary integration constant. We recognize that
this is nothing but the usual entropy of mixing formula.   An important
observation is that $S$ does not depend on $h$ explicitly and only on $M$, which is
consistent with the fact that the adiabatic lines are $M = \rm const$.  
This $M$-only dependence also satisfies the consistent condition, 
Eq.~(\ref{eqconsistent}). A
similar calculation can be carried out, and one finds
\begin{equation}
Q_2 = T_2 \Bigl(S(M_2) - S(M_1) \Bigr),
\end{equation}
thus
\begin{equation}
{ Q_2 \over Q_1} = { T_2 \over T_1}.
\end{equation}
And the desired $f$ function is
\begin{equation} 
T = f(\theta) = 1/(k_B \beta) = J/\bigl(k_B \tanh^{-1}(\theta)\bigr).
\end{equation}
The last step in this argument is to define entropy according to Clausius, i.e.,
\begin{equation}
dS =  { \delta Q \over T},
\end{equation}
or $S = -\int^M h\, dM/T + {\rm const}$.  As the integrand is a total
differential, exactly what process or which path to use is immaterial --- we get
exactly the same expression (e.g., integrating over isothermal curve).   Of
course, the result is just the standard entropy formula for paramagnet,
Eq.~(\ref{Eq(B)}).   By differentiating with respect to $E$, fixing $h$ (since
we have the fundamental thermodynamic relation $dE = TdS - Mdh$), we can see
that temperature can take both positive as well as negative values (when
$E=-hM>0$).

\subsection{\label{seccampisi}Discussion}
Boltzmann's original definition of entropy is based on probability.    
Since each state is equally
likely in a microcanonical ensemble, we have $S_B = k_B \ln W$, where $W=1/P$
is the number of microstates consistent with the given constraint, and $P$ is the 
probability of each micro-state.  For our magnetic system, 
\begin{equation}
W= {N! \over  [(N+M)/2]! [(N-M)/2]!}.   
\end{equation}
After using the Stirling approximation for the factorial function, valid for
large $N$, we obtained the standard entropy mixing formula.  

Campisi in Ref.~\onlinecite{campisi} defined a Boltzmann entropy as 
$S(E,h) = k_B \ln (W/h) + {\rm const}$.  This is not correct, according to most textbooks
(see Nagle, \cite{nagle}, who cited many popular statistical mechanics books).
Replacing the number of states by the density of states
is simply not possible as for the Ising model the energy spectrum
does not become dense even for large sizes.   If this formula is used for
entropy, we can not recover the third law of thermodynamics, as even in the ground
state $W=1$, $S$ still depends on the magnetic field $h$ logarithmically.   
Thus, the prediction by
Campisi that the thermodynamic magnetization calculated by
\begin{equation}
M = {\partial S /\partial h  \over \partial S /\partial  E } = 
-{ E \over h} - {k_B T \over h},
\end{equation}
has a singularity (the last term above) is simply due to a wrong 
application of the Boltzmann principle.  We also note that $W$, thus $S_B$,
is an even function of the magnetization $M$ only, and $M$ is an adiabatic invariant.  
But 
\begin{eqnarray}
\Omega(E,h) &=& \sum_{ - hm \leq E} 
{ N! \over [(N-m)/2]! [(N+m)/2]!} \nonumber \\
&=&\Omega^{+}(M)\theta(h) + \Omega^{-}(M)\theta(-h)
\end{eqnarray}
($m = -hN, -h(N\!-\!2), \cdots, hN$; $\theta(h)$ Heaviside step function, and
$\Omega^{+} + \Omega^{-} = 2^N$)
depends on $M$ and $h$ separately and experiences a discontinuous jump
when $h$ changes sign.   Moreover, $\omega(E,h) = \partial \Omega(E,h)/
\partial E$ is not an integrating factor, since
\begin{equation}
\label{eqdOmega}
d \Omega = \omega \delta Q + (\Omega^+ - \Omega^-) \delta(h) dh,
\end{equation}
where the heat is $\delta Q = - h dM$.
This prevents $\Omega$ being an adiabatic invariant
\cite{penrose-counter-eg}.   The appearance of the second term
in Eq.(\ref{eqdOmega}) also causes the consistent condition, 
Eq.(\ref{eqconsistent}) with $A=h$, violated.

If we reflect back on our derivation, it is clear we must have a negative temperature scale of an intensive quantity, since we have decided to measure the 
temperature, be it positive or negative, by a local property.    Boltzmann's
critical insight of connecting thermodynamic entropy with probability and chance
is throwing out of window if Gibbs volume entropy is adopted.

\begin{figure}
\includegraphics[width=\columnwidth]{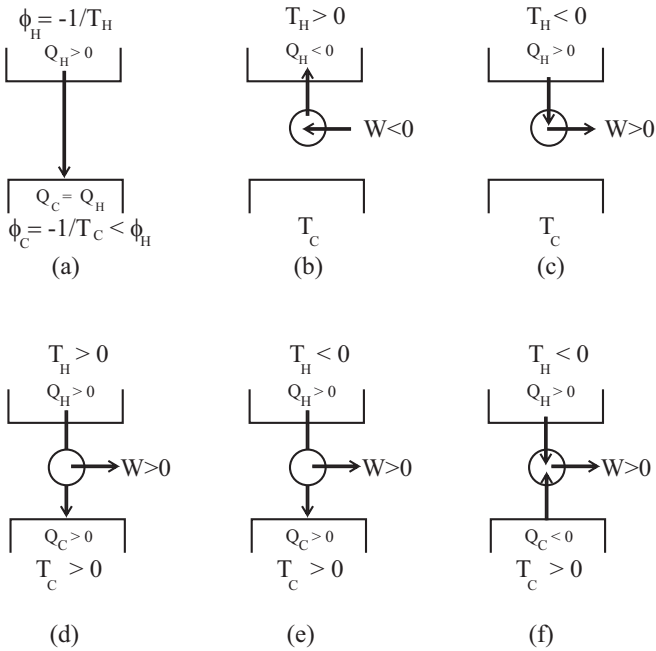}%
\caption{\label{figcarnot-cycle}Various Carnot cycle configurations with
positive and negative temperatures.}
\end{figure}


\section{\label{carnot}Carnot efficiency greater than 1}
The efficiency of a Carnot cycle when negative temperature baths are possible
is well analyzed by Ramsey, Frenkel and Warren, see also Ref.~\onlinecite{landsberg}.  If we define the Carnot
efficiency in the usual way, 
\begin{equation}
\eta = { W \over Q_H} = 1 - \frac{T_C}{T_H},
\end{equation}
where subcript $H$ means `hot' and $C$ `cold', the efficiency $\eta$ does get
values larger than one.  As explained in Ref.~\onlinecite{frenkel}, this is just a matter of
definition.  

In Figure~\ref{figcarnot-cycle}, we illustration what is possible, what is not
possible, if Clausius's second law need to be obeyed, i.e., it is not possible
for heat flowing from cold bath to hot bath spontaneously without any other
effect.  This is the only condition that can show very logically certain
processes are not possible.  Alternatively and equivalently, we can argue that
the total entropy of the combined system and the baths cannot decrease.   Let
us use the following sign convention: if the engine performs work to the
environment, $W>0$; if the engine absorbs heat from high temperature source,
$Q_H >0$; and if it releases heat, $Q_C > 0$.  energy conservation requires $Q_H
= Q_C + W$.  Then the entropy increase from the hot bath is $\Delta S_H = -Q_H
/T_H$, and from cold bath $\Delta S_C = Q_C /T_C$.  
The system/engine returns to its starting point and does not incurr
entropy change.
These equations are valid
for both positive and negative Boltzmann temperatures.  Then $\Delta S_H +
\Delta S_C \ge 0$ for adiabatic processes.  For the case of
Fig.~\ref{figcarnot-cycle}(a), $Q_H = Q_C$, we find the total entropy change need
to be $(1/T_C-1/T_H)Q_H  = (\phi_H-\phi_C)Q_H\ge 0$, here we define $\phi =
-1/T$.  If we use $\phi$ as the temperature scale, then the hotness or coldness
corresponds to the algebraic value $\phi$ being large or small.  Since we
assume $Q_H> 0$, we always need $\phi_H \ge \phi_C$, i.e., heat flows from hot
to cold bath naturally. 

We can convert work to heat with 100 per cent efficiency for positive
temperature bath (Fig.~\ref{figcarnot-cycle}(b)) but the reverse process is
forbidden by thermodynamics since that corresponds to a decrease in entropy,
$\Delta S_H = -Q_H /T_H$, if $Q_H > 0$.  However, if $T_H<0$, exactly the
opposite is true (see Fig.~\ref{figcarnot-cycle}(c)).   This last case indeed
shows that if a negative temperature bath does exist with infinite heat
capacity, then a perpetual machine of the second kind is possible.  But it is
not possible to have a sustained negative temperature bath experimentally. 

Figure~\ref{figcarnot-cycle}(d) shows the normal case of positive temperature
baths with the efficiency bounded by $0 \leq \eta < 1$, since one must release
heat, $Q_C > 0$, in order to make sure the total entropy does not decrease.  If
bath temperatures are opposite, case (f), we can use both baths to do work, 
given an efficiency
larger than one, if we still define efficiency as $W/Q_H$.  The case (e) is a
situation that is  allowed but is not a reversible process.

%
%

\section{\label{adiabat-classical}Adiabatic Invariance}
Since a strong point in promoting the Gibbs volume entropy is its adiabatic
invariance \cite{DH,campisi,berdichevsky}, 
let us try to understand what is adiabatic invariant.  S.-K. Ma, Chap 23 is helpful \cite{skma}.  In an adiabatic process we change model parameters very slowly so that the system is always in equilibrium - the important statement is we change according to
\begin{equation}
\label{eq-adb}
dE = \left\langle { \partial H \over \partial L } \right\rangle dL,
\end{equation}
using Ma's notation.  This gives the adiabatic curves in model parameter space,
$(E,L)$.  The average is over microcanonical ensemble. He proved the adiabatic
theorem using $\delta$-function notation and summing over $s$ notation which is
understandable., i.e., (set $k_B = 1$)
\begin{equation}
S(E,L) = \ln\Omega,\quad \Omega(E,L) =  \sum_{s} \theta\bigl(E-H(s,L)\bigr)
\end{equation}
is a constant if $dE = \left\langle { \partial H / \partial L } \right\rangle dL$. 

Consider the standard microcanonical ensemble for classical system defined by the
probability density in phase space as
$\rho = {\rm const}$ if $E < H < E + \Delta$ and 0 otherwise, as usual, but
keep $\Delta$ a finite quantity.   And define
\begin{equation}
\Omega(E) = \int_{H < E} d\Gamma.
\end{equation}
$d\Gamma$ is the phase space volume element for classical system.  The question
is, if $dE$ and $dL$ are related adiabatically as in Eq.~(\ref{eq-adb}), where the average
$\langle \cdots \rangle$ means over $\rho$, is $S_B = \ln \bigl(\Omega(E+\Delta) - \Omega(E)\bigr)$ a
constant?

\subsection{Simple Harmonic Oscillator Example}
Take 
\begin{equation}
H = \frac{1}{2} p^2 + \frac{1}{2} \omega^2 q^2,
\end{equation}
then
\begin{equation}
\Omega(E) = \int_{ p^2 + \omega^2 q^2 \leq 2E}\!\!\!\!\!\!\!\!\! dp\, dq = {\rm area\ of\ ellipse} = 
{2 \pi E \over \omega}.
\end{equation}
Thus, 
\begin{equation}
S_G = \ln \Omega(E) = \ln \frac{2\pi E}{\omega}
\end{equation}
and 
\begin{equation}
S_B = \ln \Bigl(\Omega(E+\Delta) - \Omega(E) \Bigr) = \ln \frac{2\pi \Delta}{\omega}.
\end{equation}
We can already see trouble here, as $S_B$ does not depend on $E$.  The differentials are
\begin{equation}
dS_G = \frac{dE}{E} - \frac{d\omega}{\omega},
\end{equation}
and 
\begin{equation}
dS_B = - \frac{d\omega}{\omega}.
\end{equation}
The adiabatic condition means, for the oscillator with a varying frequency and
energy (work done by varying frequency equals to the energy increase),
\begin{equation}
dE =  \left\langle { \partial H \over \partial \omega} \right\rangle d\omega.
\end{equation}
We need to compute the ensemble average explicitly, which is 
\begin{eqnarray}
\left\langle { \partial H \over \partial \omega} \right\rangle 
&=& \lim_{\Delta \to 0} {4 \int\int_{{\rm 1st\ quadrad}} dqdp\, \omega q^2 \over \Omega(E+\Delta) - 
\Omega(E) } \nonumber \\
&=& \frac{E}{\omega}.
\end{eqnarray}
If we maintain a finite $\Delta$, the expression is complicated, but taking the
limit makes it very simple.  This also means for the proof of adiabatic
invariance, we really need $\Delta \to 0$ limit. Now one sees that $dE = (E/\omega) d\omega$, so $S_G$ is an adiabatic
invariant, while $S_B$ is not.

This adiabaticity is closely connected to the `heat theorem' or `orthodicity'
as discussed in Gallavotti's book \cite{gallavotti}, Chap.~2.  The two are just slight
reformulate of the same question.  Gallavotti explicitly stated that in
canonical ensemble one does not need to take $N$ to infinite, but one does need
thermodynamic limit in micro-canonical ensemble (page 63-65).  

Although entropy must be an adiabatic invariant, due to Clausius equation
$dS = \delta Q/T$, there is no reason to believe that this is the only property of
entropy.  Adiabatic invariants do exist for small mechanical systems, 
in the sense demonstrated above, but these
systems do not obey zero-th law of thermodynamics.

\subsection{\label{SecQadiabaticinvarant}Adiabatic invariance in quantum systems, Penrose's counter-argument}

In quantum mechanics, one can prove an adiabatic theorem which says if one
changes the system very slowly through some model parameters, then if
the system is in a pure quantum eigenstate, it will stay in this state.
Thus, when control parameters change, if energy changes following the
instantaneous eigenvalue of the system, it is an adiabatic change.   This purely
mechanical adiabaticity coincides with thermodynamic adiabaticity, i.e., no
heat transfer.

\begin{figure}
\includegraphics[width=\columnwidth]{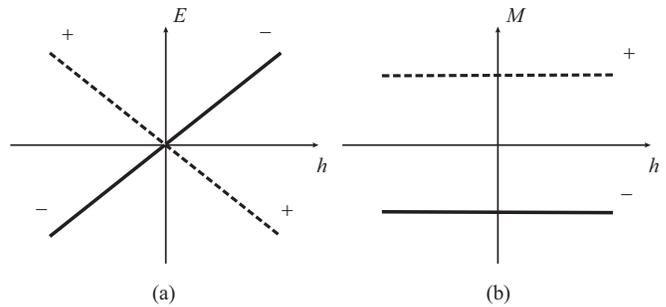}%
\caption{\label{figadiabatic}Adiabatic change in a single spin paramagnet. 
(a) The energy eigenvalues of the $+$ and $-$ states with magnetic field $h$;
(b) and associated magnetization eigenvalues.}
\end{figure}

Penrose \cite{penrose-counter-eg} gives a counter-example showing that the
total number of states $\Omega$ equal to or smaller than a given energy eigenvalue
$E_n$ is not a true adiabatic invariant.  The reason is in fact quite simple,
if there is another energy level crossing \cite{allahverdyan}
 the current energy $E_n$ which we
focus on, the number of energy levels $E_m \leq E_n$ changes for fixed $n$. 

Consider a paramagnet Hamiltonian viewed as quantum operator,
\begin{equation}
\hat H  = -h \hat{M} =  - h \sum_{i=1}^N \hat \sigma_i^z,
\end{equation}
where $\hat \sigma_i^z$ is the $z$-component of Pauli matrices.  Since $\hat
\sigma_i^z$ commutes with $\hat H$, the $z$ component of the spin is a constant
of motion from the Heisenberg equation, $i\hbar d {\hat \sigma}_i^z/dt = [\hat
\sigma_i^z,\hat H]$, for any time dependence of $h \to h(t)$.  In
Fig.~\ref{figadiabatic} we demonstrate the situation for a single spin.  If we
stay in the ground state when $h<0$, we have no other states below it, so
$\Omega=1$.  As we cross $h=0$ to the positive field region, $\Omega$ jumps
discontinuously to $2$.  Thus it is not an adiabatic invariant.  On the other
hand, the magnetization $M$ as well as counting of levels determined by $M$ are
(quantum-mechanical) adiabatic invariants.  The fact that the total
magnetization is an adiabatic invariant is fully consistent with earlier
discussion in Sec.~\ref{para-adiabats} of the Carnot cycle of paramagnet.

One might suspect that a transition occurs from `$-$' state to `$+$' state
if one goes from negative $h$ to positive $h$ slowly, such as in the Landau-Zener
problem.  But since there is no couplings between up and down spin states, this does not happen. 
If the system is in a pure quantum eigen state $|\phi_n\rangle$ of $\hat M$, 
its dynamic evolution is a trivial phase change, $\exp\bigl( \frac{i}{\hbar} M_n \int^t h(t')dt'\bigr) 
|\phi_n(0)\rangle$. Since the Hamiltonian is diagonal in the representation when
$\hat M$ is diagonal, it cannot make a transition to another state.  One can solve the von Neumann equation exactly, and finds that the density matrix evolves
according to 
\begin{equation}
\label{eqrhot}
\rho_{mn}(t) = \rho_{mn}(0) e^{\frac{i}{\hbar} (M_m-M_n) \int^t_0 h(t')dt'} ,
\end{equation}
in the representation for which $\hat M$ is diagonal.  Again we find that the
diagonal part of the density matrix does not depend on time, and off-diagonals
are oscillatory.   This dynamic evolution of the density matrix tells us that
if the system is initially in a microcanonical ensemble at $h<0$, it cannot 
change into a new and different microcanonical ensemble at $h=0$ occupying all 
micro-states with equal probability. 

It is clear if we start with a microcanonical ensemble at $(-E,-h)$, we
can evolve the system to $(E,h)$ dynamically.  But is this an adiabatic process?
To answer this question we need to 
ask what is heat and what is work, in such a driven process when
$h(t)$ varies for an initially equilibrium density matrix $\hat \rho$.  The internal energy of the system is $E = {\rm Tr}(\hat H \hat \rho)$, the differential is
\begin{eqnarray}
dE &= & {\rm Tr}(\hat H d\hat \rho) - {\rm Tr} (\hat M \hat\rho) dh \nonumber \\
\label{eqadiabatic} & = & \delta Q - M dh.
\end{eqnarray} 
We interpret the first term associated with distribution change as heat and second term work.
If we say that the change of density matrix is solely from the 
von Neumann dynamics, $d \hat\rho \propto [\hat H, \hat \rho]$, then the
first term is always 0. A correct interpretation of change must be
from re-equilibration of the system at the end of the dynamic change. 
Since the system is already in equilibrium as given by the microcanonical ensemble
(except at the point $E=h=0$), there is no need to reequilibrate, and 
we find that the heat $\delta Q$ is always 0.
Then the adiabatic curves are given by $dE = -Mdh$ \cite{adbremark}. 

Penrose used a symmetry argument to show negative temperature must exist in
such systems.   First of all, the thermodynamic entropy $S$ is a state function.  For a
paramagnet, a unique thermodynamic state is specified by $E$ and $h$, we have
$S=S(E,h)$.  Since $M=-E/h$ and $M$ is an adiabatic invariant, it means the
adiabatic curves are given by $E/h = {\rm const}$ in the state space.  In
particular, for each pair $(E,h)$, the state $(-E,-h)$ is on the same adiabatic
curve and can be reached from each other adiabatically and reversibly, so entropy
must be the same,
\begin{equation}
S(E,h) = S(-E,-h).
\end{equation}
Taking the derivative with respect to $E$, one find
\begin{equation}
\frac{1}{T(E,h)} = -\frac{1}{T(-E,-h)},
\end{equation}
i.e., the temperatures of two thermodynamic states are related by an opposite
sign.
   
The above consideration is a powerful one and very general, and certainly
not limited to a simple paramagnet.   Given any arbitrary quantum Hamiltonian
$\hat H$, with the only requirement that it has discrete eigen spectrum, we can
construct a family of Hamiltonians, $\hat H_\lambda = \lambda \hat H$.   
We see that from Eq.~(\ref{eqadiabatic}) that we can connect 
$\hat H_{+} = \hat H$ to $\hat H_{-} = - \hat H$ adiabatically, with an arbitrary process $\lambda(t)$ that varies
from $+1$ to $-1$.  This is true for any diagonal initial density matrix 
$\hat \rho$, since the diagonal elements do not evolve.   Then we must
have
\begin{equation}
\label{eq-S-equal}
S_{+}(E) = S_{-}(-E).
\end{equation} 
This equation implies if we have a positive temperature for a system with Hamiltonian $\hat H$ at energy $E = {\rm Tr}(\hat \rho \hat H)$, there must exist another system with Hamiltonian $-\hat H$ with a negative temperature at $-E$.
If the spectrum of $\hat H$ has inversion symmetry, then both positive and negative
temperature can exist in the same system, such as the Ising model.

The equality of entropies, Eq.~(\ref{eq-S-equal}), and the symmetry
argument are presented by Schneider et al \cite{schneider}, using canonical
distribution and von Neumann entropy expression.  But as we can see an explicit ensemble and entropy definition are not needed for the argument.


\section{\label{secIsing}Thermal equilibrium of an Ising model}

Consider a two-dimensional 
Ising model, which has a bounded spectrum.  For energies $E>0$, $T_B<0$, but DH
claim that the correct thermodynamic temperature is $T_G>0$.

Consider the following two experiments:

\subsection{Two Ising models with negative temperatures}

If $0>T_{B,1} > T_{B,2}$ when the systems are isolated from each other, the
temperatures will change when they are brought into thermal contact.  After
coming to a new equilibrium, we will have $0 > T_{B,1} > T_F > T_{B,2}$.

If we compare this to an experiment with the same two systems at $\vert T_{B,1}
\vert$ and $\vert T_{B,2} \vert$, we can see from symmetry that the new final
equilibrium temperature will be $\vert T_F \vert$.

No paradoxes occur.

\subsection{An Ising model and simple harmonic oscillators (SHO)}

Now consider an Ising model at $T_{B,I}   <0$ and a set of quantum simple
harmonic oscillators.  The system of SHO's must have $T_{\rm SHO} > 0$.
For simplicity, let $\hbar \omega = 2J$.  This would make it easy to simulate
the model.

If $S_B$ is correct, the temperature of the Ising system will become positive
(higher $\beta_I > 0$) if the two systems are brought into thermal contact, so
that they can exchange energy.  At the same time, $\beta_{\rm SHO}$ would decrease
(higher temperature).

DH claim that the true thermodynamic temperature of the Ising model is $T_{G,I}
> 0$.  Test their claim by setting the temperature of the SHO's to $T_{\rm SHO} =
T_{G,I}$

If DH are correct, the Ising system and the SHO's are at the same temperature,
so they should remain in equilibrium.  This is not going to happen.  It is both
obvious and can be checked by a simple simulation.

\section{\label{sec-toy}An interesting toy model}

It has occurred to us that we have available a large set of bounded energy
models.  In Monte Carlo (MC) simulations, it is standard to exclude the momenta, since they
can be integrated out exactly.

Consider a gas with only configurational degrees of freedom.
\begin{equation}
H = \sum_{j>k}
V(\vec{r}_j  - \vec{r}_k ),
\end{equation}
where
\begin{equation}
V(\vec{r})
=
X  \exp \bigl[  -\vert \vec{r} \vert^2 / (2 \sigma^2)
\bigr],
\end{equation}
and $X$ and $\sigma$ are constants.

For low positive temperature $T_B$ (large $\beta_B$), the particles will form a
crystal.

For $T = \infty$, it will be an ideal gas.

For large  $\vert T_B \vert$, with $T_B < 0$, the particles will lie
preferentially on top of each other.  This is a state that cannot be realized
for positive temperature, although DH claim that it should have a positive
thermodynamic temperature $T_{G,r}$.

Now prepare the momentum variables in an equilibrium state with temperature
$T_{G,p} = T_{G,r}$.  According to DH, allowing interactions between the
momenta and positions should not disturb equilibrium since the kinetic and
potential degrees of freedom are already at the same temperature, and therefore
in thermal equilibrium.

It is obvious that the DH claim is wrong, but it would also be easy to
simulate.

\section{\label{one-degree}One degree of freedom isolated system, or small systems}

DH are in trouble when they discuss small systems, especially a single-particle
system in one dimension.


\subsection{Two small classical systems in equilibrium}

Consider two such systems, each of which is a 1D single-particle ideal gas, in thermal equilibrium with each other, but isolated
from the rest of the world.  The probability distribution of the energy $E_1$
of the first system is 
\begin{equation} 
P(E_1) = \frac{1}{\pi
\sqrt{E_1(E_T -E_1)} }. 
\end{equation} 
This has divergent maxima at $E_1=0$ and $E_1=E_T$.
To demand equipartition seems strange.


\subsection{Specific heat of a quantum SHO}

If we've read their paper correctly, DH claim that the specific heat of a $d=1$
quantum SHO is $C=k_B$.  [Three equations after their Eq.(10), but without a
number.] This claim  is bizarre.  It would be true for a classical SHO, but all
quantum systems must satisfy the third law, which requires that $C \rightarrow
0$ as $T \rightarrow 0$.  The correct expression can be found in any textbook.

\begin{figure}
\includegraphics[width=\columnwidth]{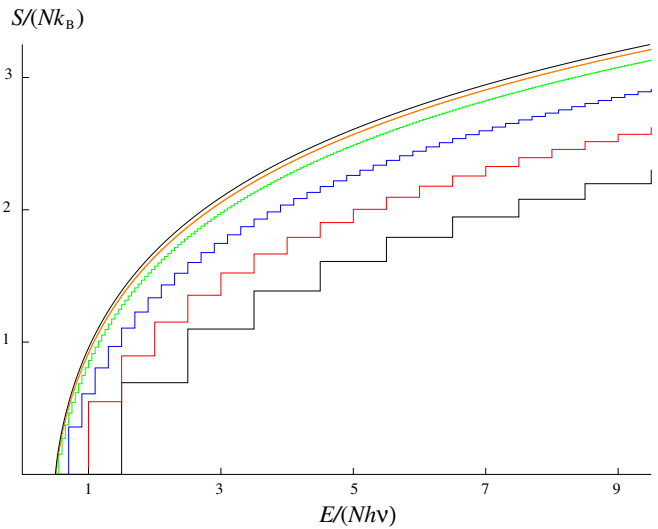}%
\caption{\label{figentropy}The entropy $S_G$ according to 
DH for harmonic oscillators with $N=1$,
2, 5, 20, 80, and $\infty$ (from bottom to top).  The limiting curve is given
by $S/(Nk_B) = (1+n)\ln(1+n) - n \ln n$, where $n = E/(Nh\nu) - 1/2$.}
\end{figure}

But this is exactly what the oscillator example shows, given a
dubious prediction that the quantum oscillator has temperature, 
\begin{equation}
\label{eq:socillatorT}
k_B T_G = \frac{ h \nu}{2} + E.  
\end{equation}
The derivation of this result is questionable \cite{derivation}.  Using their
definition of $\Omega(E) = {\rm Tr}\{\theta(E-\hat{H})\}$, the function is a
series of steps, equally spaced both in the independent variable $E$ as well as
the dependent variable $\Omega$.  Taking the logarithm to get $ S_G(E)= k_B \ln
\Omega(E)$ also results in steps, see Fig.~\ref{figentropy}.  Since the step function is a
constant for all $E$ except for a set of measure 0 which are the eigenspectrum
of the oscillator, the derivatives are 0 almost everywhere.  Thus the
temperature $T_G$ is infinite, almost everywhere, just like when the Boltzmann
entropy is applied.  In passing we also note that $S_G(E)$ violated part of
Callen's postulate III, as it is not a `continuous and differentiable'
function.

To get out of this mess, one has to consider systems with $N = 1, 2, 3,
\cdots$, oscillators in the macrocanonical ensemble, not just $N=1$.  One plots
$S/Nk_B$ vs $E/N$, as in Fig.~\ref{figentropy}.  On this scale, the steps become smaller and
smaller as $N$ becomes larger and larger, and the curves become smoother and
smoother.  In the $N \to \infty$ limit, the limiting curve is a continuous
function, and derivatives start to make sense.  That is, temperature only make
sense in the thermodynamic limit.

%
%
%
%

\section{\label{Sgproblem}Random remarks on $S_G$ and $T_G$}

Contrary to the authors' claims, it is $ S_G$ and $T_G$ that is inconsistent
with thermodynamics.  Firstly, let us consider a thought experiment with the
harmonic oscillator. We can weakly couple it with a heat bath of macroscopic in
size, then it is clear the oscillator will be in a mixed state described by the
canonical distribution.  Now we reduce the temperature $T$ of the bath to zero.
The oscillator will be cooled down to a temperature of absolute zero and stays in
its ground state.  Suppose now we turn off the coupling between the oscillator
and the bath, then there is good reason to believe that the oscillator will
still be in ground state if we turn off the coupling very slowly (quantum
adiabatic theorem).  Since the system is now isolated and the state is in
$n=0$, it is in a microcanonical ensemble with $E=h\nu/2$.  According to
Eq.~(\ref{eq:socillatorT}), then, the temperature of the oscillator becomes $h\nu/k_B$.  It
appears then, by a mere change of the mind of the experimentalist that the
system was in canonical ensemble and now in a microcanoical ensemble, the
temperature of the oscillator has been changed dramatically.   The
experimentalist can do the single oscillator experiment with CO molecule which
has a vibrational frequency of $2170\,{\rm cm}^{-1}$.  Then the associated
temperature is $T_G =  3122\,$K. The temperature of the molecule has jumped
from $0\,$K to a thousand kelvin just by changing the point of view of his
system.  One could argue that we are talking about two different temperatures,
$T_B$ and $T_G$.  But thermodynamics has only one kind of temperature.
     
Another example, consider putting a hydrogen atom in a box, in its ground state
it will be extremely hot with a $T_G$ of order $10^5\,$K, while it is cooler in
the excited states.  Another excellent example is given by contacting a finite
spin chain at (the traditional) negative $T$ with an ideal gas, as discussed by
Frenkel and Warren \cite{frenkel}.  A 3rd example is that heat
flows from cold to hot, according to the definition of $T_G$ (see Vilar and
Rubi, Ref.~\onlinecite{vilar}).   Heat flow from hot to cold naturally is one
form (Clausius) of a statement of the second law of thermodynamics.  If this is
not obeyed, it simply means that the definition or quantitative measure of the
hotness or coldness is wrong, or perhaps thermodynamics cannot apply for
such systems.   We also observe that heat flowing from hot to cold is an
immediate consequence of the Callen's second and third postulates.  This is
discussed in Callen's book, Sec.~2-4 and 2-5.  It is clear these discussions
apply equally well when temperature is negative. 

\begin{figure}
\includegraphics[width=\columnwidth]{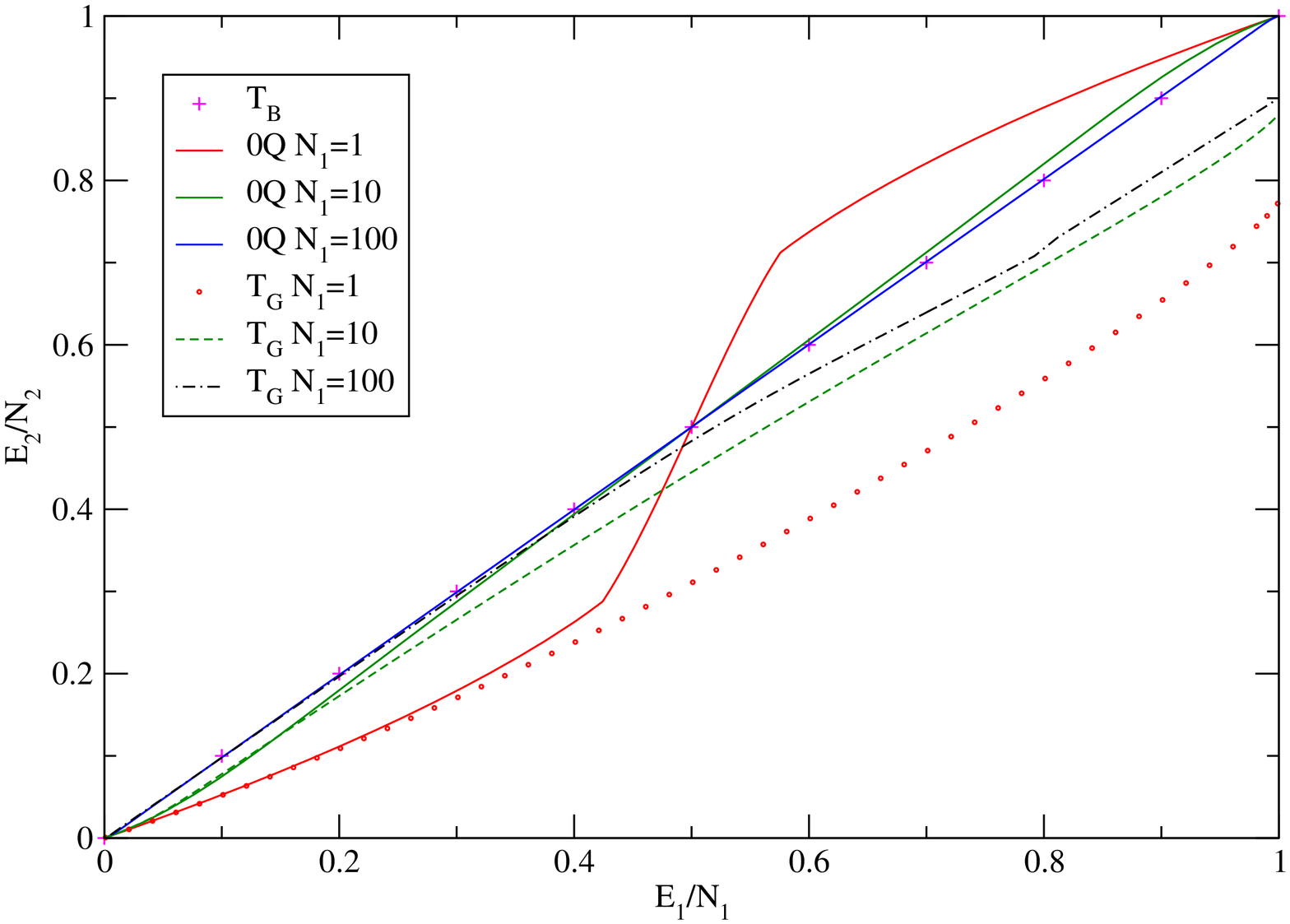}%
\caption{\label{figHHD7}Continuous energy two-level model. The lines of energy 
per particle $E_1/N_1$ vs. 
$E_2/N_2$ of two subsystems determined by equality of Boltzmann temperatures, labelled $T_B$, and 
equality of Gibbs temperatures, labelled $T_G$, as well as the lines determined by
average energies of the combined systems (lines of zero heat transfer), labelled 0Q,
with $N_1 = 1$, 10, 100, and $N_2 = 2N_1$.}
\end{figure}

In HHD (Ref.~\onlinecite{HHD}) it has been argued forcefully that neither
Gibbs temperature nor Boltzmann temperature can determine the direction of heat flow.  This is certainly true for tiny systems as several examples show, such
as that indicated by Fig.~7 in Ref.~\onlinecite{HHD}.  However, as soon as
the sizes of the systems become larger, we see clearly distinguished role that
Boltzmann temperature can play to determine the direction of heat flow while
Gibbs temperature can not, for systems with bounded energy spectra.
In Fig.~\ref{figHHD7}, we present a similar plot to the figure 7 in Ref.~\onlinecite{HHD} but with one additional parameter, the size $N_j$, $j=1$, 2.  Here we consider
a two-level system with `continuous energy' such that the Boltzmann entropy
is given by (set $k_B = 1$)
\begin{equation}
S_j^B = N_j \ln N_j  - E_j \ln E_j - (N_j - E_j) \ln (N_j - E_j),
\end{equation}
with $0\leq E_j \leq N_j,\; j = 1, 2$.
Thus, the density of states is given by $\omega_j = \exp(S_j^B)$, and the
integrated one $\Omega(E_j, N_j) = \int_0^{E_j} \omega_j(E') dE'$.  
The lines marked 0Q are the lines with no energy exchange on average 
if system 1 and 2, having energies $E_1$, $E_2$ with total energy $E_1+ E_2$ fixed, 
are combined and allowed to exchange energy.   
The 0Q lines approach rapidly the line determined by the 
equality of the Boltzmann temperatures, $T_1(E_1, N_1) = T_2(E_2, N_2)$ (which is 
given on this plot as a universal curve $E_1/N_1 = E_2/N_2$)).  The 
0Q lines do
not approach the lines determined by equality of Gibbs temperatures in the
region when the Boltzmann temperatures are negative.  Thus, thermodynamics
applicable only to large systems is not ``dogmatic'', but is really forced upon us,
as small systems do not obey thermodynamic laws. 

The use of Gibbs entropy and temperature gives a meaningless result in the
traditional negative temperature thermodynamic states,  e.g., take a 1D Ising
chain in a microcanonical
ensemble.  Then, according to DH, $S_G/N = k_B \ln 2$, and $T_G = \infty$, in
the thermodynamic limit, for all $E > 0$;
the thermodynamic heat is then $T dS = \infty \cdot 0$
which does not seem to be a useful concept.

We can elaborate a little more for the last point.  Take two segments of
sufficiently long ferromagnetic Ising chains with periodic boundary conditions,
one with a complete antiferromagnetic spin configuration, $+-+-+-+-\cdots+-$,
corresponding to the highest possible energy, the second one with some fraction
of the spins flipped over, forming $++$ or $--$ pairs, such as
$+-++-+-+--+-\cdots$.  In terms of domain walls of $+$ and $-$ boundaries, the
equal sign pairs form elementary excitations of low energy. A microcanonical
Monte Carlo dynamics can be given by flipping spins whose neighbors are
opposite in signs, which moves the domain walls around.  If we put these
segments together, it is clear that the elementary excitations will spread out
to the whole combined system to maximize the Boltzmann entropy.   But according
to DH's definition of $S_G$, this should not preferably happen, as the system
before and after combination has the same entropy per spin, $k_B \ln 2$.

The above example shows that the Gibbs entropy violated Carath\'eodory's
formulation of second law, which says one cannot have an entropy function that
is a constant in any neighborhood of an equilibrium state.  It also invalidates
Einstein theory for fluctuation.  The probability of certain amount of
fluctuation, according to Einstein, is proportional to $e^{\Delta S/k_B}$, so
macroscopic fluctuations with entropy difference of order $N$ is extremely
rare.  But if entropy is a constant, then the fluctuations of the type of
configurations discussed above is very common, which is clearly not so.  One
can verify this just by putting the system on computer and simulate. 

We now consider the last example of computer experiments.  Take again the 1D
ferromagnetic Ising chain of $N$ spins with periodic boundary condition and a 
dynamics which conserves energy.   We consider thermodynamic states and their
relationship with respect to adiabatic changes.  Here the `adiabaticity' is
in the sense of Lieb and Yngvason \cite{Lieb}, which does not implies it has 
to be slow or gentle.  We begin with an equilibrium state $X$ and end with another
equilibrium state $Y$ with the sole effect that work is done to or by the system.
The `entropy principle' states that entropy of the system must not decrease,
$S(X) \leq S(Y)$.

We consider three thermodynamic states labelled by the energies $E=-JN$,
the ground state (F), the random state $E=0$ (R), and the anti-ferromagnetically
ordered state $E=+JN$  (A).  The Gibbs entropies are 
$S_G(-JN) = \ln 2$, $S_G(0) = \ln\bigl( 2^{N-1} + N!/((N/2)!)^2\bigr) \approx
N \ln 2$ (assuming $N/2$ even), and $S_G(+JN) = N \ln 2$, and accordingly
the Boltzmann entropies are $S_B(-JN) = S_B(+JN) = \ln 2$, and $S_B(0)
= \ln \bigl( N!/((N/2)!)^2\bigr) \approx N \ln 2$.  We have the relation
$S_G(-JN) < S_G(0) < S_G(+JN)$, and $S_B(-JN) = S_B(+JN) < S_B(0)$.  
We now ask the following questions.  Can we bring state F or A to R adiabatically,
or can we bring R to F or A adiabatically?  By applying the entropy principle,
we can then judge the relative magnitude of entropies of these states.

To perform an adiabatic change, we modulate the spin coupling $J$ with a 
time-dependent protocol, $J(t)$.  Such changes do work to the system but no
heat is transferred from environment to the system, since the individual degrees of freedom $\sigma_i$  are not
manipulated.  Specifically, we turn $J$ off gradually (or otherwise), and wait
sufficiently long for the system to reequilibrate at the coupling $J=0$, and 
then turn it back to $J$ suddenly.  If we start from F or A, it ends in R.  Thus
we must have $S(F) \leq S(R)$, $S(A) \leq S(R)$.   Can we take the state R
and change it adiabatically to F or A?  If we use the above protocol $J(t)$, we
see it is very unlikely if not impossible if $N$ is finite.   But in the thermodynamic
limit R will remains in R, i.e., F to R or A to R is adiabatic irreversible, so
$S(A) < S(R)$.  This contradicts the assignment of highest possible Gibbs
entropy to state A as thermodynamic entropy. 

%

\section{\label{sec-ensemble-eq}Ensemble equivalence means thermodynamics}

The authors maintain that ensembles are generally not equivalent (particularly
if negative $T$ occurs for systems of bounded energies).  If so, there would be
two sets of thermodynamicses.  Clearly, in the traditional sense of
thermodynamics, we have only one unique one.  So in order to be consistent with
thermodynamics, different ensembles have to be equivalent \cite{touchette}.  
This happens in the thermodynamic limit.


\section{\label{sec-additive}Additivity of entropy vs positivity of temperature}

The authors gave up one fundamental property of thermodynamic systems, that is,
the extensivity and additivity (since the thermostatistical equations are supposed
to be valid for any size $N$).  An example of two-state identical boson system
(is there such a thing in nature?) was shown by the authors that $S_G$ is not
additive. Why it is reasonable to insist on positivity of $T$ but to break the
additivity?  Which property is more fundamental in thermodynamics?   If we
insist that additivity of entropy and other extensive thermodynamic variables
must be maintained, then a single quantum oscillator system, or a particle in a
box, for that matter, cannot be viewed as thermodynamic systems, as there is
nothing there to add.  The additivity is clearly needed in order to fulfill the
zero-th law of thermodynamics.

%
%

\section{\label{secCV}concavity and super-additivity}
HHD proved
\begin{equation}
\label{eq:superadd}
S^{1,2}_G(E^{1,2}) \ge S^1_G(E^1) + S^2_G(E^2).
\end{equation}
This equation shows that Gibbs entropy is super-addivity and if the system 1
and 2 refer to the same system (but can be difference sizes); it is equivalent
to concavity.  The energies are conserved, $E^{1,2} = E^1 +E^2$.  The 
assumption that internal energy is additive makes sense for macroscopic objects,
but not realistic for microscopic systems where the interaction is strong
comparing to the energies themselves saparately.

We note, firstly, that Eq.~(\ref{eq:superadd}) that HHD has proved
is generally a greater than sign `$>$' not greater or equal `$\ge$' sign.
This is so for most systems where the density of states are strictly positive
for positive energies, such as an ideal gas, and also so for most of the
parameters if energy is bounded as well. This means putting two systems together
always leads to a strict increase of Gibbs entropy, and 
splitting up causes a decrease of entropy.   Secondly, this equation is
not what Planck has in mind.  Planck has been always thinking in terms of two
systems and making them in thermal contact and then putting them apart (see,
Sec.122-126 of Planck's ``Treatise on Thermodynamics'' \cite{planckbook}).  So
the proper way to write Planck's expression of the second law should be
\begin{equation}
S'_1 + S'_2 \ge S_1 + S_2,
\end{equation}
when the system changed from the non-prime states to the primed states
adiabatically.  Of course, Boltzmann entropy also has the same strictly `$>$'
sign, but if thermodynamic limit is taken (as it can be identified with the
thermodynamic entropy only when the size is large), then we will not have
problem.

Consider quantum two-level system (equivalent to a paramagnet) with energies 0
and $\epsilon$, respectively, of distinguishable particles.    $H =
\sum_{i=1}^N n_i \epsilon$, $n_i = 0, 1$, $\epsilon > 0$.

System 1 at highest energy $E^1 = N\epsilon$, the Boltzmann entropy is,
\begin{equation}
S^1_B = \ln 1 = 0,
\end{equation}
and Gibbs
\begin{equation} 
S^1_G = \ln 2^N, 
\end{equation}
taken $k_B = 1$. 

Second system 2 at the second highest energy $E^2 = (N-1)\epsilon$, one
particle must be in the lower energy state, the rest $N-1$ in high energy state.  So 
\begin{equation}
S^2_B = \ln N,
\end{equation}
and Gibbs
\begin{equation} 
S^2_G = \ln (2^N-1). 
\end{equation}
If we combine the two systems to get $(1,2)$, the energy is still the second
highest, $E^{1,2} = (2N-1)\epsilon$.  So the entropies of combined systems are
\begin{equation}
S^{1,2}_B = \ln (2N),
\end{equation}
and Gibbs
\begin{equation} 
S^{1,2}_G = \ln (2^{(2N)}-1). 
\end{equation}
The entropy increases, are
\begin{equation}
\Delta S_B = S^{1,2}_B - S^1_B - S^2_B= \ln 2 > 0,
\end{equation}
and Gibbs
\begin{equation} 
\Delta S_G = S^{1,2}_G - S^1_G - S^2_G = \ln (1 + 1/2^N) >0. 
\end{equation}
So for both definitions, entropies increase, so ``second law'' is obeyed for
both.  But the actual numerical values are very different -- Boltzmann entropy
is a finite amount of 0.69, the Gibbs entropy extremely small for a large
system.  The temperatures are also very different  (using approximation $T
\approx \delta E / \Delta S$), $T_B \approx -\epsilon/\ln 2 = O(1)$, negative;
$T_G =2^N \epsilon \to \infty$.  Which one makes thermodynamic sense?

%

\section{\label{sec-postulate-II}Callen's postulate II}
Concavity is a weaker condition than being entropy postulated by Callen, as we
cannot see why concavity implies Callen's postulate II.  
This postulate in
terms of math, means
\begin{equation}
\label{eq-callenII}
S^{1,2}(E) = \max_{E^1} \Bigl( S^1(E^1) + S^2(E-E^1) \Bigr),
\end{equation}
$E = E^1 + E^2$ is the total energy which is constrained between system 1 and
2, which is also the total  energy of the combined system, (1,2).  The value
$E^1$ for which the right-hand side obtains its maximum, $E_{\rm max}^1$, is
the value system 1 will taken when the constraint is removed.

\subsection{\label{SecNtest}A numerical test}
We use the same non-interacting two-level system model as in Sec.~\ref{secCV},
i.e., for a single particle the energy is 0 in ground state and $\epsilon$ in
excited state.  We set $\epsilon=1$ for numerical convenience.  The entropies
are
\begin{equation}
S_B(N,E) = \ln { N! \over E!\, (N-E)!}, \quad 0 \leq E \leq N,
\end{equation}
and 
\begin{equation}
S_G(N,E) = \ln \sum_{k=0}^E{ N! \over k! \,(N-k)!}.
\end{equation}
Since $\epsilon = 1$, the energy $E$ also takes integer values.

Consider two systems, system 1 with particle number $N_1$, and second system 2 of
identical system except the number of particles is $N_2 = 2N_1$, twice larger.
We set the energy of the combined system $(1,2)$ to $E^{1,2} = \frac{4}{5}(N_1 +
N_2)$.   This will make the combined system with negative $T_B$.  We ask, for
what value of $E^1$, the right-hand side of Eq.~(\ref{eq-callenII}) is a
maximum?  This table shows the numerical answers:
\vskip 5mm
\begin{tabular}{|r|r|r|r|}
\hline\hline
\emph{size $N_1$} & \emph{$E^1_{\rm max}$ from $S_B$} &
 \emph{$E^1_{\rm max}$ from $S_G$} & \emph{\% error} \\
\hline
5 	&	4	&	4 	&	0\% \\
10	&	8	&	9 	&	11\% \\
50 	&	40	&	43	&	7.5\%	\\
100	&	80	&	87	&	8.7\%	\\
500	&	400	&	433	&	8.2\%	\\
1000	&	800	&	867&	8.4\%	\\
\hline\hline
\end{tabular}
\vskip 5mm
\noindent Clearly, when the two systems are combined with a particle number
$3N_1$, the excitation energies should be distributed evenly to the original
two systems to maximising entropy, i.e., it should give $E_{\rm max}^1 =
(4/5)N_1$, at least for large enough systems.   The Boltzmann entropy predicted
this down to small sizes 5 correctly, while Gibbs entropy always gives more
energy to system 1 and less to system 2.  The right-most column in the
table gives the percentage errors made in the energy.  The deviations do not
seem to decrease as $N_1$ goes to infinite (thus it is not a $1/N$ effect).  
Thus, we have demonstrated
numerically that $S_G$ violated Callen's postulate II.  

The locations of the maxima are associated with the equality of 
respective temperatures.  An exact calculation can be carried out for the
equilibrium values of energy of system 1 to be
$\langle E^1 \rangle = \frac{N_1}{N_1 + N_2} E^{1,2}$ in the combined
system.  The heat exchange between the two subsystems,
$Q = \langle E^1 \rangle - E_{\rm max}^1 = 0$ if we start at the energies
predicted by the Boltzmann entropy being largest, and Boltzmann temperatures
being equal.
 
But from HHD point of view, this violation of Callen's postulate II is 
irrelevant as they rejected Callen's formulation of thermodynamics.

\section{\label{sec-HHD-ThermoD}HHD thermodynamics}

HHD proposed a thermodynamics which we express as the following:
\begin{quotation}
Zeroth law: temperature of the parts is the same as the temperature of the 
whole when in thermal contact.  
Note that it is forbidden to ask what the temperatures will be if the parts are 
separated.  In fact, the parts get different Gibbs temperatures if one does so. But thermodynamics alone cannot predict them.

First law: there is no dispute on this; we have energy conservation, 
$dE = \delta Q + \delta W$,
$\delta Q = T\,dS$.  However, even here, $T$ and $S$ are not state functions.

Second law: entropy strictly increase (generically), i.e., $S^{1,2} > 
S^1 + S^2$, when two bodies are combined.  How much it increase?
thermodynamics alone cannot tell.
\end{quotation}

%
%

The above formulation of thermodynamics may well be correct for
small systems, but there is very little prediction power it can offer. 
There is not much one can do with it.  For example, the formulation has no 
way to tell us which direction heat flows if two bodies are made in thermal contact.  There is no numerical relation of the entropy of the combined system $(1,2)$ with that  of separated systems 1 and 2, uncoupled.  One cannot
analyze the efficiency of a Carnot engine, as the formulation does not have
provision for the thermal coupling/decoupling of the system and the baths.

The way 
to try to understand why thermodynamics has to be formulated in such a way
is that the authors trying to fit the bill of Gibbs $S_G$.  The Gibbs entropy
does satisfy all the properties listed above \cite{0thlaw}.  The `thermodynamic laws' are
invented to `check' that $S_G$ satisfies them.  Clearly, the argument
is logically circular; and that $S_G$ satisfies the thermodynamic laws
is a misleading claim.    

One may argue that we are misrepresenting things, as the above formulation
is for isolated systems.  Even so, why it is useful at all?

%

\section{\label{sec-split-join}Nonequilibrium Statistical-Mechanical Analysis of the coupling/decoupling processes}

We give an analysis as what happen if we couple and decouple two identical equilibrium systems, from statistical mechanics point of view and see if entropy increase or decrease.

a) Consider two identical boxes of ideal gases of macroscopic sizes with exactly
the same macroscopic measurable properties, i.e., the same thermodynamic state.  
If there is a partition separating them that can be open or closed, does the total entropy change?   Of course not, according to the usual thermodynamics.  In fact
this is one of the fundamental axiom in Lieb-Yngvason's formulation of 
thermodynamics (Ref.~\onlinecite{Lieb}, page 21, axiom (A5), splitting and 
recombination).  The entropy 
of macroscopic observables does give zero entropy change by definition
\cite{swendsen-wang2014,RHS1,RHS book,RHS unnormalized,RHS Change
2014}.
This contradicts the HHD second law.

b) Now we consider systems as small statistical mechanical systems described
by either a classical Hamiltonian dynamics or quantum dynamics for the distributions, $\rho$.  
Open or closing of a partition can be modeled by
a time-dependent external potential.  First of all, if we start with a product of microcanical distributions, $\rho_1 \rho_2$, of two separate systems (let's call it statistical-mechanical state $P$), it is not possible to evolve into 
a new microcanonical distribution of combined system (call it $C$), simply Liouville's theorem
prevents the value of probability $\rho$ to be changed along the dynamic trajectory. 
Also when separating the system, it is also not possible to evolved back to a
product state, $P$.  The von Neumann entropy is a constant in any events.
So the question about entropy increase or decrease cannot be answered in such a
framework.

In order to have a Boltzmann-like H-theorem, one need to use a coarse-grained
description \cite{vonNeumann,lebowitz,penrose,RHS1}, which essentially
lead to:

c) a third option, a stochastic dynamics of Markov chain.  In this framework the problem is well-posed.
In such a description, we can make transitions from $P$ to $C$ dynamically 
which lead to an entropy increase.  Still $C$ to $P$ is not possible.
If we start from the combined system $C$ in a microcanonical distribution,
the probability distribution ends up to a distribution of energy partitions of
the product microcanonical distributions, let us call this state $D$, the decoupled
but still entangled state, so $C \to D$.   

To think of this with a concrete example, we can still take the paramagnet
or equivalently the two-level system, with energy $\epsilon$ in the excited
state and 0 for the ground for each particle.  To give a meaning to spatial
locality and partition, we imagine that the spins/particles form a 1D chain.  The
dynamics is simply to pick two nearest neighbor particles at random 
and swap the locations.
This is an energy (and magnetization) conserving process. Let us consider two identical systems
with $N$ particles each and one excited particle each with total energy $E=\epsilon$.  The entropy
increases when two systems are combined are, 
$\Delta S_B =  S_B(C) -S_B(P) = \ln (2-1/N)$, and 
$\Delta S_G = \ln \bigl[ (2N^2+N+1)/(N+1)^2\bigr]$, respectively.  For both definitions
of entropy, the increase approaches $\ln 2$ for sufficiently large $N$.

If we split the system from state $C$ by adding partition, it evolves towards $D$.  
It is not clear at all how to quantify the entropy.  We give four possibilities: (i) we know
exactly what the energies are of each subsystems --- such is the case if we simulate the process on computer, and check exactly how many excited particles are in one of the
partitions.  Then total entropy decreases.  (ii) If we are not allowed to measure
the energy or our measurement is not precise, the total entropy still decreases. For
example, the total entropy may be defined by $S_{{\rm ii}}(D) = \sum_{j} p_j (S^1_j + S^2_j)$,
where $p_j$ is the probability that subsystem 1 get an energy $E^1_j$ which can be
calculated from the initial microcanonical distribution of system $C$.  Then, (iii), we can
apply the Shannon entropy, $- \sum \rho \ln \rho = -\langle \ln \rho \rangle$, 
over the whole state space.   Since the probability of a particular
microscopic configuration is the probability $p_j$ of getting split into energy $E^1_j$,
times the probability in a particular microscopic state of the joint system, which is equal probable in each subsystem, we can write,
\begin{equation}
S_{{\rm iii}}(D) = S_{B,{\rm ii}}(D) + \bigl\{ - \sum_{j} p_j \ln p_j \bigr\}.
\end{equation}
This is higher comparing to case (ii) by an amount due to the random selection of energies.
For the two particles in a 1D box of size $2N$ problem, one finds that 
$S_{{\rm iii}}(D) - S_B(C) = 0$ (this is a general result), and 
$S_{{\rm iii}}(D) - S_G(C) < 0$, i.e., entropy does not change if Boltzmann entropy for 
state $C$ is used and entropy still decreases if Gibbs entropy is used. 
Curiously, $C$ to $D$ is a reversible process or recoverable - removing the parition will
make the system back to $C$, so entropy does not change make good sense.
(iv) In the last case we can consider the expression given by \cite{tasaki,joshi}
$\sum_i \rho_i \ln \Omega(E_i) = 
\langle \ln \Omega \rangle$, where $E_i$ is the energy of the system
in micro-state $i$ with probability $\rho_i$.   With the microcanonical constraint,
all states have the same energy, so it is a constant with the value
$S_G(C)$; entropy does not change.  Unfortunately, it is a constant no matter what initial distributions to
begin with.  So, it cannot describe relaxation towards equilibrium.

To conclude, if we believe that the opening or closing a partition is an adiabatic
process (possibly performing work but no heat exchange), then entropy must not decrease.
Cases (i)-(iv) show that Gibbs entropy contradicts Planck's formulation of the second law,
or at least, Gibbs entropy cannot make any definite statement as it is not properly defined for state $D$. The fact that we cannot have unambiguous interpretation of entropy in a
state  $D$ only demonstrates that thermodynamic entropy is a macroscopic concept.  The 
contradictions and ambiguity disappear for large systems. 
   
%
%

\section{\label{sec-non-mono}Non-monotonic dependence of $T$ with energy $E$}
HHD cited reasons not to use temperature $T_G$ or $T_B$ to judge if heat can
flow from one body to another or not.  Both of them fail to do so 
for tiny systems. The reason
for this failure is because they insists that thermodynamics works for any size
$N$.  If one consider only thermodynamic limit result (in the sense of omitting
any contributions smaller than $N$ for extensive quantities such as energy
and entropy, and omitting quantities of order $1/N$ or $\log( N)/N$ for
intensive quantities such as temperature), thermodynamics works just fine.  Both
zeroth and second law can be satisfied which define the thermodynamic system. 

The internal energy $E$ is a monotonic function of $-\beta$  for large systems,
so increasing temperature always lead to an increase of the
internal energy.  This can be seen easily in the canonical ensemble, since in
the canonical ensemble the heat capacity measures the fluctuation of the
energy, 
\begin{equation}
C = \frac{dE}{dT} = \frac{1}{k_B T^2} \bigl[ \langle H^2 \rangle - \langle H \rangle^2 \bigr],
\end{equation}
which is always positive, and one can verify that the heat capacity is
always a positive quantity, even for the negative Boltzmann temperature states.

%

\section{\label{sec-why}Why insists on $S_G$?}
By postulate $S_G$ as the thermodynamic entropy, we got 1) positivity of $T$
(but why this is important), 2) concavity and super-additivity of $S_G$.  But
breaks many things:  1') ensembles no longer equivalent, even for a paramagnet,
2') temperatures of individual systems and combined system are not the same,
and violating the usual 0-th law of thermodynamics.  3') Violation of Callen's
postulate II.  4') meaningless thermodynamics when $T_B<0$, which are $T_G =
\infty$, $S_G = {\rm const}$, heat capacity $C=0$.  One cannot write down a
meaningful thermodynamic relation as it is $\infty \cdot 0$. 5') thermodynamics
losses its predicting power in the sense of Callen's postulate II since $S_G$
is a constant.  In addition, when applied to single quantum degree systems (we
believe one cannot do that for both $S_B$ and $S_G$), one violates 6') third
law of thermodynamics.

HHD's deduction has a logic fallacy:  thermodynamic entropy must satisfy
zeroth, first, and second law.  Gibbs volume entropy does not satisfy HHD's version
of zeroth law (average temperatures of subsystems when in thermal contact equal the temperature as a whole), when energy of the system is bounded.  
Therefore, there is no logical implication that thermodynamic temperature 
cannot be negative.    

%

\section{\label{sec-final}Final remark}
If canonical ensemble were used to treat the harmonic oscillator, an explicit
thermodynamic limit is not necessary as the free energy is already a
homogeneous function of particle number even for one and two particles.  It is
clear we are already in the``thermodynamic limit''; the specific one calculated
for one particle in a canonical ensemble is just one of the many.  However, in
the microcanonical ensemble, situation is different and taking the limit is a
necessary step in order to have an entropy expression that is consistent with
thermodynamics. 

\section*{Acknowledgements}
The author thanks Robert H. Swendsen for many useful exchanges and
even contributions to some of the passages in this critique.  But the opinions
are the author only. He also
thanks Peter H\"anggi for an inspiring talk, although we disagree on most
of the issues discussed here.  Finally, I sincerely apologize to both of them
that this work has caused unnecessary damage on an emotional level and 
damage to our long-lasting friendship.   Each one of us holds stubbornly a
view unreconcilable with the other, which may be a good thing for physics.


\begin{thebibliography}{01}

\bibitem{purcell} E. M. Purcell and R. V. Pound, ``{\sl
A nuclear spin system at negative temperature,}'' Phys. Rev. 
\textbf{81}, 278 (1951).

\bibitem{ramsey} N. F. Ramsey, ``{\sl Thermodynamics and statistical mechanics at negative absolute temperatures,}'' Phys. Rev. \textbf{103}, 20 (1956).

\bibitem{schreider-exp} S. Braun, J.P. Ronzheimer, M. Schreiber, S. S.  Hodgman, T. Rom, I. Bloch, U. Schreider, ``{\sl Negative absolute temperature for motional degrees of freedom,}'' Science \textbf{339}, 52 (2013).

\bibitem{DH} J. Dunkel and S. Hilbert, ``{\sl Consistent thermostatistics forbids negative absolute temperatures,}'' Nature Phys. \textbf{10}, 67 (2014).

\bibitem{HHD}
S. Hilbert, P. H\"anggi, and J. Dunkel,
``{\sl Thermodynamic laws in isolated systems,}''
Phys. Rev. E {\bf 90}, 062116 (2014).

\bibitem{DHreply} J. Dunkel and S. Hilbert, ``{\sl Reply to Frenkel and Warren,}'' arXiv:1403.6058.

\bibitem{DH reply to Schneider}
J. Dunkel and S. Hilbert,
``{\sl Reply to Schneider et al,}''
arXiv:1408.5392v1, (2014).

\bibitem{campisi} M. Campisi, ``{\sl Construction of microcanonical entropy
on thermodynamic pillars},'' Phys. Rev. E \textbf{91}, 052147 (2015).

\bibitem{schneider} U. Schneider, S. Mandt, A. Rapp, S. Braum, H. Weimer, I. Bloch, and A. Rosch, ``{\sl Comment on `consistent thermostatistics forbids negative absolute temperatures,}'' arXiv:1407:4127.


\bibitem{vilar} J. M. G. Vilar and J. M. Rubi, ``{\sl Communication: System-size scaling of Boltzmann and alternate Gibbs entropies,}'' J. Chem. Phys. \textbf{140}, 201101 (2014).

\bibitem{frenkel} D. Frenkel and P. B. Warren, `{\sl Gibbs, Boltzmann, and negative temperatures,}'' Am. J. Phys. \textbf{83}, 163 (2015).

\bibitem{swendsen-wang2014} R. H. Swendsen and J.-S. Wang, ``{\sl Negative temperatures and the definition of entropy,}'' arXiv:1410.4619; Am. J. Phys. to appear. 

\bibitem{swendsen-wang2015} R. H. Swendsen and J.-S. Wang, ``{\sl
The Gibbs `volume' entropy is incorrect,}'' arXiv:1506.066911.

\bibitem{penrose-counter-eg} O. Penrose, ``Counter-argument to
`Thermostatistical consistency forbids negative absolute temperature' by
J. Dunkel and S. Hilbert (Nature Phys. \textbf{10}, 67-72 (2014)),''  unpublished.

\bibitem{buonsante} P. Buonsate, R. Franzosi, and A. Smerzi,
``{\sl Phase transitions at high energy vindicate negative microcanonical temperature,}'', arXiv:1506.01933.

\bibitem{Nlimit} We do not need to take $N$ literately to infinity.  It only means for mathematical expressions of quantities such as entropy,  terms smaller than $O(N)$ can be dropped.

\bibitem{callen} H. B. Callen, ``{\sl Thermodynamics and an Introduction to Thermostatistics,}'' 2nd ed (John Wiley \& Sons, New York, 1985).

\bibitem{Giles} R. Giles,  ``{\sl Mathematical foundations of thermodynamics,}''
Pergamon, Oxford (1964).

\bibitem{Lieb} E. L. Lieb and J. Yngvason, ``{\sl The physics and mathematics
of the second law of thermodynamics,}'' Phys. Rep. 310, 1-96 (1999).

\bibitem{pippard} A. B. Pippard, ``{\sl The Elements of Classical Thermodynamics,}''  (Cambridge Univ. Press, Cambridge, 1957).

\bibitem{romero} V. Romero-Rochin, ``{\sl Nonexistence of equilibrium states at absolute negative temperatures,}'' Phys. Rev. E \textbf{88}, 022144 (2013).

\bibitem{RHS1} R. H. Swendsen,
 ``Statistical mechanics of classical systems with distinguishable particles,'' 
 J. Stat. Phys., \textbf{107}, 1143--1165  (2002).

\bibitem{RHS book} R.H. Swendsen,
\textit{An Introduction to Statistical Mechanics and Thermodynamics}, 
(Oxford, London, 2012).

\bibitem{RHS unnormalized}  R.H. Swendsen,
``Unnormalized Probability:
A Different View of Statistical Mechanics,''
Am. J. Phys., \textbf{82}, 941 (2014).                     

\bibitem{RHS Change 2014}  R.H. Swendsen,
``Entropy is a Description of Macroscopic Change,''
unpublished.

\bibitem{gallavotti} G. Gallavotti, ``{\sl Statistical Mechanics: A short treatise,}''  (Springer, Berlin, 2010). 

\bibitem{skma} S.-K. Ma, ``{\sl Statistical Mechanics,}'' (World Scientific, Singapore, 1985).

\bibitem{penrose} O. Penrose, ``{\sl Foundations of Statistical Mechanics,
a deductive treatment,}'' Dover ed. (2005). 

\bibitem{nagle} J. F. Nagle,  ``{\sl In defense of Gibbs and the traditional
definition of the entropy of distinguishable particles,}'' Entropy, 
{\textbf 12}, 1936 (2010).

\bibitem{landsberg} P. T. Landsberg, ``{\sl Thermodynamics and 
Statistical Mechanics,}'' Chap.~10, Oxford Univ. Press (1978).

\bibitem{berdichevsky} V. L. Berdichevsky, ``{\sl Thermodynamics of 
Chaos and Order,}'' (Addison-Wesley/Longman, Harlow, 1997).

\bibitem{allahverdyan} A. E. Allahverdyan and Th. M. Nieuwenhuizen,
``{\sl Minimal work principle: proof and counterexamples,}''
Phys. Rev. E \textbf{71}, 046107 (2005).

\bibitem{adbremark} We can interpret the change of density matrix
$\hat \rho$ in three ways, 1) quantum dynamics of the von Neumann
equation with finite speed, 2) infinite slow adiabatic change such that
the state remains in the instantaneous eigen state of $H(t)$, 3) thermalization
in each step of the way with equilibrium $\rho_{{\rm eq}}$.  Fortunately, for 
the present problem, they all give the same adiabats, $E/h = {\rm const}$.
Particularly noteworthy is that at the point $(E,h)=(0,0)$, the system cannot
be in thermal equilibrium, but it does not matter - there is no heat entering 
or leaving the system, or generated internally, as long as we do not stop at
$(0,0)$; thus the process cannot be quasi-static but still is adiabatic.  Alternative
view (Penrose): since $M$ is a conserved quantity, the microcanonical
ensemble should be characterized by both $E$ and $M$.  Then quasi-static
process is also possible passing the point $(E,h)=(0,0)$. 

\bibitem{derivation} DH derived $T_G$ in the following way.  Consider a quantum 
system with energy eigenvalues $E_n$, $n=1,\,2,\,3,\,\cdots$, such that $n=1$ is 
the ground state.  Then, by identifying $E$ with $E_n$ and $\Omega$ with $n$, 
we get $k_B T_G = n\, dE/dn$.  For harmonic oscillator, $E = (n-1/2)h\nu$, 
and for a particle in a box, $E=bn^2$, we obtain Eq.~(\ref{eq:socillatorT}) for the oscillator 
and $k_B T_G = 2E$ for a particle in a box.  These results violated third law 
of thermodynamics -- temperature is not zero when entropy is zero.  
Another example will be more instructive: consider a bounded 1D potential 
such that the energy spectrum is hydrogen-like, $E_n = - R/n^2$, $R>0$.  
Then $k_B T_G = 2R/n^2 = -2E$.  We obtain this result that the 
ground state is hotter than the excited states.

\bibitem{touchette} For a recent proof, see, H. Touchette,
``Equivalence and nonequivalence of ensembles: Thermodynamic, macrostate, and measure levels,'' arXiv:1403.6608.

\bibitem{planckbook} M. Planck, ``{\sl Treatise on Thermodynamics,}'' 
Dover ed. (1945).

\bibitem{0thlaw} Gibbs entropy does not satisfy HHD's version of 0-th law for
systems with bounded energies, see Fig.~6 in \cite{HHD}. In contrast to this,
if hotness or coldness is numerically measured in terms of $\beta = 1/(k_B T)$,
the Boltzmann $\beta_B$ satisfies their 0th law.   

\bibitem{vonNeumann} J. von Neumann, ``{\sl Beweis des Ergodensatzes und des 
H-Theorems in der neuen Mechanik,}'' Zeitschrift f\''{u}r Physick, {\bf 57}, 30-70 (1929);
English translation by R. Rumulka, arXiv:1003.2133.

\bibitem{lebowitz} J. L. Lebowitz, ``{\sl From time-symmetric microscopic dynamics
to time-asymmetric macroscopic behavior: an overview,}'' in
G. Gallavotti, W. L. Reiter, and J. Yngvason (editors), 
{\sl Boltzmann's Legacy}, Euro. Math Soc. (2008);  arXiv:0709.0724.

\bibitem{tasaki}
H. Tasaki, arXiv:cond-mat/0009206 (2000).

\bibitem{joshi} D. G. Joshi and M. Campisi,
``{Quantum Hertz entropy increase in a quenched spin chain,}''
Eur. Phys. J. B 86, 1 (2013).

\end{thebibliography}
\end{document}